\documentclass[12pt,nohyper,notoc]{JHEP}

\usepackage{amsmath,euscript,array,amssymb,cite} 

\newcommand{\startappendix}{
\setcounter{section}{0}
\renewcommand{\thesection}{\Alph{section}}}
\newcommand{\Appendix}[1]{
\refstepcounter{section}
\begin{flushleft}
{\large\bf Appendix \thesection: #1}
\end{flushleft}}

\def\ed{\end{document}}
\def\aD{{\dot\alpha}}
\def\bD{{\dot\beta}}
\def\CM{{{\cal M}}}
\def\N{{\cal N}}

\def\sst{\scriptscriptstyle}
\def\det{{\rm det}}
\def\SU{\text{SU}}
\def\U{\text{U}}

\def\Sp{\text{Sp}}

\newcommand{\BL}{\boldsymbol{L}}

\newcommand{\Bomega}{{\boldsymbol{\omega}}}

\newcommand{\Bchi}{{\boldsymbol{\chi}}}
\newcommand{\Bphi}{{\boldsymbol{\phi}}}
\def\Dbarslash{\,\,{\raise.15ex\hbox{/}\mkern-12mu {\bar\D}}}
\def\Dslash{\,\,{\raise.15ex\hbox{/}\mkern-12mu \D}}
\def\delslash{\,\,{\raise.15ex\hbox{/}\mkern-9mu \partial}}
\def\delbarslash{\,\,{\raise.15ex\hbox{/}\mkern-9mu {\bar\partial}}}
\def\K{{\cal K}}

\def\ms{{\mathfrak M}}
\def\ns{{\mathfrak N}}
\def\F{{\mathfrak F}}

\def\Q{{\cal Q}}

\newcommand{\PD}[2]{\frac{\partial #1}{\partial #2}}
\newcommand{\MAT}[1]{\begin{pmatrix} #1\end{pmatrix}}
\newcommand{\EQ}[1]{\begin{equation} #1 \end{equation}}
\newcommand{\AL}[1]{\begin{subequations}\begin{align} #1
\end{align}\end{subequations}}
\newcommand{\ALT}[2]{\begin{subequations}\begin{alignat}{#1} #2
\end{alignat}\end{subequations}}
\newcommand{\SP}[1]{\begin{equation}\begin{split} #1 \end{split}\end{equation}}

\title{Calculating the Prepotential by Localization on the Moduli
Space of Instantons}

\author{Timothy J.~Hollowood\\
Department of Physics, University of Wales Swansea,
Swansea, SA2 8PP, UK\\
E-mail: {\tt t.hollowood@swan.ac.uk}
}

\abstract{We describe a new technique for calculating instanton
effects in supersymmetric gauge theories applicable on the Higgs or
Coulomb branches. In these situations the instantons are constrained and a
potential is generated on the instanton moduli space. Due to
existence of a nilpotent fermionic symmetry 
the resulting integral over the instanton moduli space
localizes on the critical points of the potential. Using this
technology we calculate the one- and two-instanton contributions to the
prepotential of $\SU(N)$ gauge theory with $\N=2$ supersymmetry and
show how the localization approach yields the prediction extracted from 
the Seiberg-Witten curve. The technique appears to extend to 
arbitrary instanton number in a tractable way.
}

\keywords{}

\preprint{{\tt hep-th/0201075}\\SWAT-328}

\begin{document}

\section{Introduction}

Instantons are particularly fascinating in the context of
supersymmetric theories. Certain physical quantities are known on the
basis of non-renormalization theorems to receive non-perturbative
instanton contributions which are semi-classically exact. For instance
in this work we will be concerned with the prepotential of an $\N=2$
supersymmetric gauge theory with---for definiteness---$\SU(N)$ gauge symmetry.
Of course, the celebrated theory of Seiberg and Witten \cite{SeibWitt}
gives an entirely different way to calculate the instanton expansion
of the prepotential, however, this does not detract from the dream of
calculating the prepotential directly from the path integral using
semi-classical techniques. For one thing
consistency shows that the Euclidean path integral does actually have
something to do with Minkowski space field theory beyond perturbation
theory and also that the
remarkable edifice of Seiberg-Witten theory is consistent with more
conventional approaches.

The problem is that instanton calculations are not easy. In the
semi-classical limit the functional integral reduces to a sum over
terms involving 
integrals over the moduli spaces of instantons $\ms_{k,N}$ of charge
$k$.\footnote{Here, $N$ keeps track of the gauge group.}
The $4kN$-dimensional moduli space $\ms_{k,N}$ can be constructed via the
Atiyah-Drinfeld-Hitchin-Manin (ADHM) construction \cite{ADHM}; however,
this realizes $\ms_{k,N}$ as a hyper-K\"ahler quotient and so the
construction is only implicit: there are complicated non-linear
constraints which have not at present been solved, for general $N$,
beyond $k=1$ (except on generic orbits for $N\geq2k$ \cite{MO3,MY}).
Even in the case of a single instanton, where the
``ADHM constraints'' can be solved, the resulting integral over
$\ms_{1,N}$ which give the one-instanton contribution to the prepotential
is rather complicated for $N>2$, as one can appreciate by following
the machinations of
Ref.~\cite{KMS}.\footnote{In fact, if one inspects the approach of
Ref.~\cite{KMS} closely one will see that the ADHM constraints are
never actually solved: they are replaced by integrals over 
Lagrange multipliers, a trick which we will use below.} 
A direct attack on the two-instanton contribution can
only be performed for the gauge group $\SU(2)$ (actually realized via
the $\Sp(1)$ ADHM construction) \cite{MO-I,MO-II}. The two-instanton
calculation was a {\it tour de force\/} in its own right and 
sadly the extension to $\SU(N)$ and/or $k>2$ looks a touch unrealistic. 
A new idea is urgently needed and this is what the present work provides.

The prepotential of an $\N=2$ theory with $N_F$ fundamental
hypermultiplets has the form\footnote{In the case of the finite theory,
obtained when $N_F=2N$, one must replace the factor $\Lambda^{2N-N_F}$
by $e^{2\pi i\tau}$ where, as usual, $\tau=4\pi i/g^2+\theta/2\pi$.}
\EQ{
{\cal F}={\cal F}_{\text{classical}}+{\cal F}_{\text{one-loop}}
+\frac1{2\pi i}\sum_{k=1}^\infty{\cal F}_k\Lambda^{k(2N-N_F)}\ .
\label{expf}
}
where $\Lambda$ is the usual scale of the Pauli-Villars regularization
scheme.\footnote{For $N_F=2N$ the factor $\Lambda^{2N-N_F}$ must by replaced by
$e^{2\pi i\tau}$.} The coefficient ${\cal F}_k$ is given by
\EQ{
{\cal F}_k=\widehat{\EuScript Z}_{k,N}^{\sst(\N=2,N_F)}\ ,
\label{ndeq}
}
the ``centred instanton partition function'' defined as 
an integral over the suitably $\N=2$ supersymmetrized
version of the centred $k$-instanton moduli space
$\widehat\ms_{k,N}$.\footnote{The centred moduli space has the overall
position of the instanton configuration factored off:
$\ms_{k,N}={\mathbb R}^4\times\widehat\ms_{k,N}$.} If
$\Bomega^{\sst(\N=2,N_F)}$ is the $\N=2$ supersymmetric volume form, then
\EQ{
\widehat{\EuScript Z}_{k,N}^{\sst(\N=2,N_F)}=\int_{\widehat\ms_{k,N}}
\Bomega^{\sst(\N=2,N_F)}\,e^{-S}\ .
\label{coeff}
}
The quantity $S$ is the instanton effective action which 
depends on the VEVs parameterizing the Coulomb branch in addition to 
the masses of hypermultiplets. This action is a direct manifestation of the 
fact that instantons are
not exact solutions of the equation-of-motion on the Coulomb branch,
rather they should be treated as constrained instantons {\it \`a la\/}
Affleck \cite{Affleck,ADS} (see also the in-depth discussion in
Refs.~\cite{MY,MO-I}). 

The partition function \eqref{coeff} can be obtained as the 
dimensional reduction to zero dimensions of the partition function of
a two-dimensional $\N=(0,4)$ supersymmetric $\sigma$-model with
$\widehat\ms_{k,N}$ as target \cite{MY}. It is possible to linearize this
$\sigma$-model by introducing a non-dynamical $\U(k)$ vector multiplet
containing a two-dimensional gauge field. Integrating out the vector
multiplet simply implements the hyper-K\"ahler quotient construction of
$\ms_{k,N}$. The VEVs of the four-dimensional gauge theory can then be
incorporated via a non-trivial dimensional reduction.
It is important to bear in mind that the original
$\SU(N)$ gauge symmetry of the four-dimensional $\N=2$ theory is realized as a
global symmetry of this zero-dimensional ``field'', or matrix, 
theory. Will find the linearized formulation of the partition function
particularly advantageous for our calculations. 

The key point, following 
Refs.~\cite{Dorey:2001zq,Fucito:2001ha},\footnote{The former 
reference considers the
analogous construction in the context of the $\N=4$ theory.} is 
that there exists a fermionic
symmetry $\Q$ defined as a particular combination of the 
supersymmetries for which the instanton effective action has the form
\EQ{
S=\Q\,\Xi+\Gamma\ ,
\label{specs}
}
where $\Q\,\Gamma=0$. 
The operator $\Q$ can be shown to be nilpotent on quantities invariant
under the $\U(k)$ auxiliary symmetry group and also
under the $\U(1)^{N-1}\subset\SU(N)$ subgroup of the global symmetries 
picked out by the VEVs of the four-dimensional $\N=2$
theory. In
particular, the action $S$ and integration measure
$\Bomega^{\sst(\N=2,N_F)}$ are $\Q$-invariant. Now consider the more 
general integral
\EQ{
\widehat{\EuScript Z}_{k,N}^{\sst(\N=2,N_F)}(s)
=\int_{\widehat\ms_{k,N}}\Bomega^{\sst(\N=2,N_F)}\,\exp\big(
-s^{-1}\Q\,\Xi-\Gamma\big)\ .
\label{suggvg}
}
We then have
\EQ{
\PD{\widehat{\EuScript Z}_{k,N}^{\sst(\N=2,N_F)}(s)
}{s}=
s^{-2}\int_{\widehat{\ms_k}}\Bomega^{\sst(\N=2,N_F)}\,\Q
\Big\{\Xi\,\exp\big(-s^{-1}\Q\,\Xi-\Gamma\big)\Big\}\ ,
\label{varet}
}
using the fact that $\Q^2\,\Xi=\Q\Gamma=0$. 
Since the volume form is $\Q$-invariant
the right-hand side of \eqref{varet} 
vanishes. Consequently, $\widehat{\EuScript Z}_{k,N}^{\sst(\N=2,N_F)}(s)$
is independent of $s$ and, therefore, it can be evaluated
in the limit $s\to0$ where the integral is dominated by
the critical points
of $\Q\,\Xi$. Since the result is independent of $s$,
under favourable circumstances---which will be shown to hold
in the present application---the Gaussian approximation is exact 
(for references to this kind of localization in the physics literature see
Refs.~\cite{berline,Niemi:1994ej,Blau:1995rs,Schwarz:1997dg} and references 
therein). In the present case, the presence of an effective action for
instantons arises from the fact that instanton are constrained on the
Coulomb branch. The action acts as a potential on $\ms_{k,N}$ which
penalizes large instantons and the critical points 
can be identified with the configurations where all the instantons
have shrunk to zero size. The potential problem is that the theory of
localization is most easy to apply to situations involving
compact spaces without
boundary. In the case at hand, the instanton moduli space is obviously
not compact and, in addition, has conical singularities precisely when
instantons shrink to zero size. On top of this,
the critical-point sets are non-compact and have their own
singularities when the
point-like instantons can becomes arbitrarily separated in ${\mathbb
R}^4$ or they can come together at the same point in ${\mathbb R}^4$,
respectively. We will argue that the potential problems 
are removed by considering the 
partition function in the $\N=2$ theory defined on a {\it
non-commutative\/} spacetime.\footnote{Actually in order to define a
non-commutative version of the theory we have to take a gauge group
$\U(N)$. The addition of the extra $\U(1)$ factor does not affect the
commutative theory since it trivially decouples.}
In such a theory,
instantons are described by a deformed moduli space
$\ms^{(\zeta)}_{k,N}$, where $\zeta^c$, $c=1,2,3$, are
non-commutativity parameters, that is a smooth 
resolution of $\ms_{k,N}$: instantons can no longer shrink to zero
size on a non-commutative spacetime.\footnote{Various aspects of
instantons in non-commutative theories are considered in 
Refs.~\cite{NS,Furuuchi123,Lee:2001hp,KLY12,Nekrasov:2000zz,Schwarz:2001ru,Chu:2001cx,Hamanaka:2001dr}.}
Of course the question then arises as to whether it is legitimate in
the context of the prepotential to replace the instanton moduli space
by its smooth resolution? It has been argued in
Ref.~\cite{Hollowood:2001ng} that this
is indeed the case and that the physical content of the 
prepotential of the non-commutative theory is the same as that
of the commutative theory.\footnote{There are differences which
turn out to be non-physical, as we shall see.}

Before proceeding we should point out that the construction of a
BRST-type operator $\Q$ in the context of the $\N=2$ instanton calculus was
first proposed in
Refs.~\cite{Fucito:2001ha,Bellisai:2000tn,Bellisai:2000bc}. (In
particular, the latter reference is most closely related to the
approach that we adopt in this paper.) However,
these references emphasized the relation with the
topological version of the original gauge theory. However, these
references did not go on to use the existence of $\Q$ to develop a
calculational technique based on localization. A nilpotent fermionic
symmetry was also constructed in the context of the $\N=4$ instanton
calculus in Ref.~\cite{Dorey:2001zq} where localization was first
proposed as a method to calculate, in this case the $\N=4$, instanton
partition function. Some recent papers \cite{Flume:2001nc} have also
considered the $\Q$-operator and the $\N=2$ instanton calculus, although
they use the equivalent language of differential forms
in which $\Q$ corresponds to
an equivariant exterior derivative on the instanton moduli
space. These references then go some way towards interpreting ${\cal
F}_k$ as a topological intersection number.

\section{The Instanton Calculus and Localization}

In this section, we will briefly review relevant aspects of the
instanton calculus, construct the fermionic symmetry and prove the
properties of the centred instanton partition function that we
described above. For a more detailed discussion of the instanton
calculus we refer the reader to Refs.\cite{MO3,MY,KMS}.

The ADHM construction of instantons involves a set of over-complete
collective coordinates $\{w_\aD,a'_n\}$. Here, $w_\aD$, $\aD=1,2$, are
$N\times k$ matrices with elements $w_{ui\aD}$ and $a'_n$,
$n=1,2,3,4$, are $k\times k$ Hermitian matrices.\footnote{$\SU(N)$
gauge indices are denoted $u,v=1,\ldots,N$ and ``instanton'' indices
are denoted $i,j=1,\ldots,k$, where $k$ is the instanton charge.} 
The instanton moduli
space $\ms_{k,N}$ is then obtained as a hyper-K\"ahler quotient
\cite{Hitchin:1987ea} by the
group $\U(k)$ acting on the variables as
\EQ{
w_\aD\to w_\aD\,{\cal U}\ ,\qquad a'_n\to{\cal U}^\dagger\,
a'_n\,{\cal U}\ ,
\label{symm}
}
${\cal U}\in\U(k)$. 
The hyper-K\"ahler quotient involves a two-stage process. First one
defines the ``level set'' 
$\ns$, to be the subspace of the ``mother''
space, in this case simply ${\mathbb R}^{4k(N+k)}$ parameterized
by $\{w_\aD,a'_n\}$, on which the $3k^2$ following ``ADHM
constraints'' are satisfied (known also as the vanishing of the 
moment maps):\footnote{Here, $\tau^c$, $c=1,2,3$, are the Pauli matrices,
$a'_{\alpha\aD}=a'_n\sigma_{\alpha\aD n}$ and $\bar
a^{\prime\aD\alpha}=a'_n\bar\sigma_n^{\aD\alpha}$, where
$\sigma_n=(i\tau^c,1_{\sst[2]\times[2]})$ and
$\bar\sigma_n=(-i\tau^c,1_{\sst[2]\times[2]})$. Spinor indices
$\aD,\bD,\ldots$ 
are raised and lowered with the $\epsilon$-tensor as in \cite{WB}. Finally, 
$\bar w_\aD\equiv(w_\aD)^\dagger$.}
\EQ{
\tau^{c\aD}{}_\bD\big(\bar w^\bD w_\aD+\bar
a^{\prime\bD\alpha}a'_{\beta\aD}\big)=\zeta^c1_{\sst[k]\times[k]}\ .
\label{adhm}
}
Here, $\zeta^c$ are real constants which are set to zero if we wish to
describe $\ms_{k,N}$, whereas they are left non-trivial,
$\zeta\cdot\zeta>0$, in order to describe the smooth resolution
$\ms_{k,N}^{(\zeta)}$. The instanton moduli
space is then the ordinary quotient of $\ns$ by the $\U(k)$ action
\eqref{symm} which fixes $\ns\subset{\mathbb R}^{4k(N+k)}$. 
In order to define the
centred moduli space $\widehat\ms_{k,N}^{(\zeta)}$ one simply imposes the fact
that $a'_n$ are traceless. The relation of the instanton calculus to
the hyper-K\"ahler quotient construction has been emphasized in
Refs.~\cite{MY,Bruzzo:2001di}. 

We remark at this point that the construction of
$\ms_{k,N}$ is valid even in the case $N=1$. In this case when
$\zeta^c=0$ it is
easy to see that the constraints \eqref{adhm} are solved with
$w_\aD=0$ and $a'_n$ diagonal; hence, taking into account that the
diagonal form for $a'-N$ is fixed by permutations, we have
\EQ{
\ms_{k,1}=\frac{{\mathbb R}^{4}\times\cdots\times{\mathbb R}^4}{S_k}\ ,
\label{sins}
}
where the symmetric group $S_k$ acts by permutation. One can show that
the resulting expression for the $\U(1)$ gauge field is pure
gauge. In the non-commutative $\U(1)$ theory, instantons become
non-trivial and $\ms_{k,1}^{(\zeta)}$ is a smooth resolution of the
singular space \eqref{sins} 
\cite{NS,Nakajima:1993jg}. For example, the centred moduli space
$\widehat\ms_{2,1}^{(\zeta)}$ is a smooth four-dimensional
hyper-K\"ahler space which can be shown to be the Eguchi-Hanson
manifold \cite{Lee:2001hp}. 

In an $\N=2$ supersymmetric theory an instanton has a set of Grassmann
collective coordinates which parameterize the $4kN$ zero modes of the
Dirac operator in the instanton background. In the supersymmetric
extension of the ADHM construction we first define a set of
over-complete Grassmann variables $\{\mu^A,\bar\mu^A,\CM^{\prime
A}_\alpha\}$. Here $A=1,2$ is an index of the $\SU(2)$
$R$-symmetry,\footnote{These indices are raised and lowered with the
usual $\epsilon$-tensor of Ref.~\cite{WB}.}
$\mu^A$ is an $N\times k$ matrix with elements $\mu^A_{ui}$,
$\bar\mu^A$ is a $k\times N$ matrix with elements $\bar\mu^A_{iu}$ and
$\CM^{\prime A}_\alpha$ are $k\times k$ matrices with elements $(\CM^{\prime
A}_\alpha)_{ij}$. This over-complete set of variables is subject to fermionic
analogues of the ADHM constraints:
\EQ{
\bar\mu^A w_\aD+\bar w_\aD\mu^A+[\CM^{\prime\alpha A},a'_{\alpha\aD}]=0\ .
\label{fadhm}
}
For the centred moduli space we also impose the tracelessness of the
matrices $\CM^{\prime A}_\alpha$. Geometrically we can identify the independent
Grassmann collective coordinates with tangent vectors to 
$\ms_{k,N}^{(\zeta)}$.

When we include $N_F$ fundamental hypermultiplets there are additional
Grassmann collective coordinates $\K$ and $\tilde\K$ which are
$k\times N_F$ and $N_F\times k$ matrices with elements $\K_{if}$ and
$\tilde\K_{fi}$, $f=1,\ldots,N_F$, 
respectively. These coordinates are not subject to
any constraints.

We now specify the volume form $\Bomega^{\sst(\N=2,N_F)}$. For
the moment, we consider the full moduli space $\ms_{k,N}^{(\zeta)}$
rather than its centred version. The integral can be split into a
$c$-number piece times various Grassmann integrals:
\EQ{
\int_{\ms_{k,N}^{(\zeta)}}\Bomega^{\sst(\N=2,N_F)}\equiv
\int_{\ms_{k,N}^{(\zeta)}}\Bomega\times\text{Grassmann integrals}\ .
\label{diffgp}
}
The $c$-number volume form $\Bomega$ is the canonical volume form on
$\ms_{k,N}^{(\zeta)}$. This can be deduced from the 
hyper-K\"ahler quotient. The quotient starts
from flat space with metric
\EQ{
ds^2=
8\pi^2{\rm tr}_k\,\big[d\bar w^\aD\,dw_\aD+da'_n\,da'_n\big]\ .
\label{qum}
}
A natural metric is then induced on the hyper-K\"ahler 
$\ms_{k,N}$ by restriction to $\ns$ and then via a conventional quotient by
$\U(k)$. If this metric is $g=g_{\mu\nu}dX^\mu\,dX^\nu$, in terms of a
set of intrinsic coordinates $\{X^\mu\}$ on $\ms_{k,N}$, then 
\EQ{
\int_{\ms_{k,N}^{(\zeta)}}\Bomega=
\int\sqrt{\det\,g}\prod_\mu\frac{dX^\mu}{\sqrt{2\pi}}\ .
}
Unfortunately, in general beyond $k=1$, 
the ADHM constraints have not been solved and so it is not
possible to find an explicit set of intrinsic coordinates on
$\ms_{k,N}^{(\zeta)}$. However, we can write down an implicit expression
which leaves the ADHM constraints explicit and where we don't fix the
$\U(k)$ symmetry \cite{measure1,DHKM}. 
One of the main conclusions from our analysis is
that we can make progress even when we {\it cannot\/} solve the ADHM
constraints. This implicit form for the collective coordinate
integration measure for the full moduli space, following
Refs.~\cite{KMS,MO3} for the $\SU(N)$ case, is\footnote{In
the following we use the following definitions. The integrals over
$a'_n$ and the arguments of the $\delta$-functions are defined with 
respect to the generators of $\U(k)$ in the fundamental
representation, normalized so that
${\rm tr}_k\,T^rT^s=\delta^{rs}$. The volume of the
$\U(k)$ is the constant 
$2^k\pi^{k(k+1)/2}/\prod_{i=1}^{k-1}i!$.}
\SP{
\int_{\ms_{k,N}^{(\zeta)}}
\Bomega&=\frac{2^{-k(k-1)/2}(2\pi)^{2kN}}{{\rm Vol}\,\U(k)}\, \int\, 
d^{2kN}w\,d^{2kN}\bar w\,d^{4k^2}a'\\
&\qquad\qquad\times 
\,\big|\det_{k^2}\BL\big|\ \prod_{r=1}^{k^2}\, \prod_{c=1}^3\,
\delta\Big(\tfrac12{\rm
tr}_k\,T^r\big(\tau^c{}^\aD{}_\bD( \bar w^\bD
w_\aD+\bar
a^{\prime\bD\alpha}a'_{\alpha\aD})-\zeta^c1_{\sst[k]\times[k]}
\big)\Big)\ .
\label{bmes}
}
Here, $\BL$ is an operator on $k\times k$ matrices defined by
\EQ{
\BL\cdot\Omega=
\tfrac12\{\bar w^\aD w_\aD,\Omega\}+[a'_n,[a'_n,\Omega]]\ .
\label{vvxx}
}
The fact that \eqref{bmes} follows from the hyper-K\"ahler quotient
construction was shown in Ref.~\cite{Bruzzo:2001di} (see also the
review \cite{MY}).
Geometrically, the matrix element ${\rm tr}_k(T^r\BL T^s)$, for two
generators $T^r,T^s$ of $\U(k)$, gives the inner-product of the corresponding
Killing vectors on the mother space of the hyper-K\"ahler quotient \cite{MY}.

There are two Grassmann pieces to the collective coordinate 
integral \eqref{diffgp}. The first
corresponds to the zero modes of the adjoint-valued fermions which
involve integrals over $\{\mu^A,\bar\mu^A,\CM^{\prime A}_\alpha\}$ 
subject to the
fermionic analogues of the ADHM constraints \eqref{fadhm}. The form of
the integral is dictated by the hyper-K\"ahler quotient
construction \cite{MY}. For $\N$ supersymmetries these integrals
are 
\cite{KMS,DHKM,MO3}\footnote{Unlike \cite{WB} out convention for integrating a
two-component spinor $\psi_\alpha$ is $\int
d^2\psi\equiv\int\,d\psi_1\,d\psi_2$.}  
\SP{
\prod_{A=1}^\N\bigg\{ 2^{-kN}\pi^{-2kN}
& \int\,d^{kN}\mu^A\,d^{kN}\bar\mu^A\,d^{2k^2}\CM^{\prime A}
\,\big\vert\det_{k^2}\,\BL\big|^{-1}\\
&\times \ \prod_{r=1}^{k^2}\, \prod_{\aD=1}^{2}\,\delta\Big({\rm
tr}_k\,T^r(
\bar\mu^Aw_\aD+\bar w_\aD\mu^A+[{\cal M}^{\prime\alpha
A},a'_{\alpha\aD}])\Big)\bigg\}\ .
\label{intsf}
}
To complete the collective coordinate integral there are integrals
over the Grassmann collective coordinates arising from the fundamental
hypermultiplets:
\EQ{
\pi^{-2kN_F}\int\,d^{kN_F}\K\,
d^{kN_F}\tilde\K\ .
\label{hint}
}

To define the centred partition function we have to separate out the
integrals over the trace parts of $a'_n$ and $\CM^{\prime A}_\alpha$. These
collective coordinates are generated by translations and global
supersymmetry transformations on the instanton configuration. Taking
into account the normalization of the associated zero modes the
prescription for defining the centred partition function is
\EQ{
\int_{\ms_{k,N}^{(\zeta)}}\Bomega^{\sst(\N=2,N_F)}=\frac4{\pi^2}\int
d^4({\rm tr}_ka')\,d^4({\rm tr}_k\CM')\ \times\ 
\int_{\widehat\ms_{k,N}^{(\zeta)}}\Bomega^{\sst(\N=2,N_F)}\ .
}

The final ingredient of the centred partition function is the instanton
effective action. This has been determined in \cite{KMS} (see the review
\cite{MY} for full details). The expression depends on the VEVs of the
scalar fields. Sometimes it will be convenient to think of the VEVs as
being a real 2-vector $\Bphi$
and sometimes as the (anti-)holomorphic quantity
$\phi$ ($\phi^\dagger$) with
\EQ{
\Bphi=({\rm Re}\,\phi,{\rm Im}\,\phi)\ .
}
The quantity
$\phi$ is then a diagonal $N\times N$ matrix with
elements $\phi_u$. The traceless
condition implies $\sum_{u=1}^N\phi_u=0$.
The expression for the effective action is
\SP{
S&=4\pi^2{\rm tr}_k\Bigg\{-\tfrac
i{2}
\bar\mu^A\phi^\dagger\mu_A+\bar
w^\aD\Bphi^2w_\aD\\
&+\Big(\tfrac 1{4}\sum_{f=1}^{N_F}
\K_f\tilde\K_f-\bar w^\aD\phi^\dagger w_\aD\Big)\BL^{-1}\Big(
-\tfrac i2\bar\mu^A\mu_A-\tfrac i{2}\CM^{\prime\alpha A}\CM'_{\alpha A}
+\bar w^\aD\phi w_\aD\Big)\Bigg\}\ .
\label{ieant}
}
Note that the action penalizes large instantons as is characteristic
of constrained instantons. 

The centred instanton partition function can be formulated in an
equivalent but more useful way by introducing auxiliary variables in
the form of: $\Bchi=(\chi_1,\chi_2)$ a 2-vector of Hermitian $k\times
k$ matrices; $D^c$, $c=1,2,3$, 
three Hermitian $k\times k$ matrices; and $k\times k$ matrices
of Grassmann superpartners $\bar\psi^\aD_A$. Using these variables, the
partition function can be written in a completely ``linearized''
form:
\SP{
\widehat{\EuScript Z}_{k,N}^{\sst(\N=2,N_F)}
&=\frac{2^{-k(k-1)/2-2}\pi^{-4k^2-2kN-2kN_F+2}}
{{\rm Vol}\,\U(k)}
\int d^{2kN}w\,d^{2kN}\bar w\,d^{4(k^2-1)}a'\,d^{3k^2}D\,
d^{2k^2}\Bchi\\
&\times\,d^{2kN}\mu\,d^{2kN}\bar\mu\,
d^{4(k^2-1)}\CM'\,d^{4k^2}\bar\psi\,d^{kN_F}\K\,d^{kN_F}\tilde\K\,\exp(-S)\ ,
\label{ipf}
}
where the action is
\SP{
S&=4\pi^2{\rm tr}_k\Big\{\big|w_\aD\Bchi+\Bphi w_\aD\big|^2
-[\Bchi,a'_n]^2
+\tfrac i2\bar\mu^A(\mu_A\chi^\dagger+\phi^\dagger\mu_A)\\
&\qquad\qquad+\tfrac i4\CM^{\prime\alpha
A}[\CM^{\prime}_{\alpha A},\chi^\dagger]+
\tfrac14\sum_{f=1}^{N_F}(m_f-\chi)\K_f\tilde\K_f\Big\}+ 
S_{\text{L.m.}}\ .
\label{jslmm}
}
Here, $\chi=\chi_1+i\chi_2$.
Notice that the $k\times k$ matrices $D^c$
and $\bar\psi^\aD_A$ act as Lagrange multipliers
for the bosonic and fermionic ADHM constraints \eqref{adhm} and
\eqref{fadhm} through the final term in the action
\EQ{
S_{\text{L.m.}}=-4i\pi^2{\rm
tr}_k\Big\{\bar\psi_A^\aD\big(\bar\mu^A 
w_\aD+\bar w_\aD\mu^A+[\CM^{\prime\alpha A},
a'_{\alpha\aD}]\big)+D^c\big(\tau^{c\aD}{}_\bD(\bar w^\bD w_\aD+\bar
a^{\prime\bD\alpha}a'_{\alpha\aD})-\zeta^c\big)\Big\}\ .
}  
The previous form of the collective coordinate integral, obtained by
concatenating \eqref{bmes}, \eqref{intsf} and \eqref{hint}, is
recovered by integrating out the auxiliary variables
$\{\Bchi,D^c,\bar\psi^\aD_A\}$. 

The instanton effective action \eqref{jslmm} (or \eqref{ieant}) 
is invariant under four supersymmetries corresponding to 
precisely half the number of the $\N=2$ theory. 
On the linearized system, the transformations can be written as
\ALT{2}{
\delta a'_{\alpha\aD}&=  i\bar{\xi}_{\dot{\alpha}A}
{\cal M}^{\prime A}_{\alpha}\ ,&\qquad
 \delta{\cal M}^{\prime A}_{\alpha}& =  2\bar\xi^{\aD A}
[a'_{\alpha\aD},\chi]\ ,\label{susy12}\\
\delta w_\aD&= i\bar\xi_{\aD A} \mu^A\ ,&\qquad
\delta \mu^A&=  2\bar\xi^{\aD A}\big(w_\aD\chi+\phi w_\aD\big)\ ,
\label{susy3}\\
\delta\chi&=0\ ,&\qquad\delta \chi^\dagger&=2i\bar{\xi}^{A}_\aD
\bar\psi_A^\aD\ ,\\
\delta \bar\psi_{A}^\aD &=
\tfrac12[\chi^\dagger,\chi]\bar\xi_{A}^\aD-iD^c\tau^{c\aD}{}_\bD 
\bar\xi_{A}^\bD\ .&\delta\K&=\delta\tilde\K=0\ .
\label{susy22}
}

One can interpret the partition function 
in the linearized form \eqref{ipf} as the dimensional reduction of a
two-dimensional 
$\N=(0,4)$ supersymmetric gauged linear $\sigma$-model \cite{MY}. In this
interpretation $\Bchi$ is the $\U(k)$ two-dimensional gauge field
forming a vector multiplet of supersymmetry along with
$\bar\psi_A^\aD$ and $D^c$. These variables have no kinetic term (in
two dimensions) and on integrating them out 
one recovers a non-linear
$\sigma$-model with the hyper-K\"ahler space 
$\widehat\ms_{k,N}^{(\zeta)}$ as target.

From the supersymmetry transformations we can define corresponding 
supercharges via
$\delta=\xi_{\aD A}Q^{\aD A}$. The fermionic symmetry we are
after can then be defined as
\EQ{
\Q=\epsilon_{\aD A}Q^{\aD A}.
}
Notice that the definition of $\Q$ mixes up spacetime and $R$-symmetry
indices as is characteristic of topological twisting. 
The action of $\Q$ on the variables is\footnote{This action is
essentially identical to the one defined in Ref.~\cite{Fucito:2001ha}
after integrating out some additional auxiliary variables. It can also
be derived, as explained in Ref.~\cite{Fucito:2001ha}, from the
$\Q$-action defined in the $\N=4$ theory in Ref.~\cite{Dorey:2001zq}.}
\ALT{2}{
\Q\, w_\aD&=i\epsilon_{\aD A}\mu^A\ ,&\qquad\Q\,\mu^A&=-2\epsilon^{\aD A}(
w_\aD\chi+\phi w_\aD)\ ,\\
\Q\, a'_{\alpha\aD}&=i\epsilon_{\aD A}\CM^{\prime A}_\alpha\ ,&\qquad
\Q\,\CM^{\prime A}_\alpha&=-2\epsilon^{\aD A}[a'_{\alpha\aD},\chi]\ ,\\
\Q\,\chi&=0\ ,&\qquad\Q\,\chi^\dagger&=2i\delta^A{}_\aD\bar\psi^\aD_A\ ,\\
\Q\,\bar\psi^\aD_A&=\tfrac 12\delta^\aD{}_A[\chi^\dagger,\chi]
-iD^c\tau^{c\aD}{}_\bD\delta^\bD{}_A\ ,&\qquad\Q\,\K&=\Q\,\tilde\K=0\ .
}
It is straightforward to show that $\Q$ is nilpotent up to an infinitesimal 
$\U(k)\times\SU(N)$ transformation generated by $\chi$ and $\phi$. 
For example,
\EQ{
\Q^2\,w_\aD=w_\aD\chi+\phi w_\aD\ .
}

It can then be shown that the instanton effective 
action assumes the form \eqref{specs} with
\SP{
\Xi&=4\pi^2{\rm tr}_k\Big\{\tfrac12\epsilon_{\aD A}\bar
w^\aD\big(\mu^A\chi^\dagger
+\phi^\dagger\mu^A\big)+\tfrac14\epsilon_{\aD A}\bar
a^{\prime\aD\alpha}[\CM^{\prime
A}_\alpha,\chi^\dagger]\\
&\qquad\qquad+\delta^{A}{}_\aD\bar\psi^\bD_A\big(\bar w^\aD
w_\bD+\bar
a^{\prime\aD\alpha}a'_{\alpha\bD}-\tfrac12\tau^{c\aD}{}_\bD\zeta^c\big)\Big\}
}
and
\EQ{
\Gamma=\pi^2\sum_{f=1}^{N_F}{\rm tr}_k\big(
(m_f-\chi)\K_f\tilde\K_f\big)\ .
}
Parenthetically we note that the derivative of the action with
respect to the variables $\zeta^c$ is $\Q$-exact. This means that the
partition function cannot depend smoothly on the non-commutativity parameters
$\zeta^c$. Of course, there will be a discontinuity at $\zeta^c=0$
when singularities appear on $\ms_{k,N}^{(\zeta)}$.

Following the logic of localization we should investigate the 
critical points of $\Q\,\Xi$. The terms to minimize are, from
\eqref{jslmm},
\EQ{
\big|w_\aD\Bchi+\Bphi w_\aD\big|^2
-[\Bchi,a'_n]^2\ .
}
Notice that this is positive semi-definite and the critical points are
simply the zeros:
\EQ{
w_\aD\Bchi+\Bphi w_\aD=[\Bchi,a'_n]=0\ .
}
Up to the $\U(k)$ auxiliary symmetry, each critical-point set is
associated to the partition
\EQ{
k\to k_1+k_2+\cdots+k_N\ .
}
Each $i\in\{1,2,\ldots,k\}$ is then 
associated to a given $u$ by a map $u_i$ as follows:
\SP{
&\Big\{\underbrace{1,2,\ldots,k_1}_{u=1},
\underbrace{k_1+1,\ldots,k_1+k_2}_{u=2},\ldots,\\
&\qquad\ldots,\underbrace{k_1+\cdots+k_{u-1}+1,
\ldots,k_1+\cdots+k_u}_u,\ldots,\ldots,
\underbrace{k_1+\cdots+k_{N-1}+1,\ldots,k}_{u=N}\Big\}
}
For a given partition the variables have a block diagonal-form
\EQ{
\Bchi_{ij}=-\Bphi_{u_i}\delta_{ij}\ ,\qquad
w_{ui\aD}\propto \delta_{uu_i}\ ,\qquad
(a'_n)_{ij}\propto\delta_{u_iu_j}\ .
\label{critp}
}
The critical-point sets have a very suggestive form. Imposing the
ADHM constraints \eqref{adhm} implies that in the $u^{\rm th}$ 
block the constraints are those of $k_u$ instantons in a
non-commutative $\U(1)$ gauge theory.
The critical-point set associated to $\{k_1,\ldots,k_N\}$ is then simply
\EQ{
{\mathfrak F}=\frac{\ms^{(\zeta)}_{k_1,1}\times
\cdots\times\ms^{(\zeta)}_{k_N,1}}{{\mathbb R}^4}\ ,
\label{gfps}
}
where the quotient is by the overall centre of the
instanton. The factors $\ms^{(\zeta)}_{k,1}$ are the non-commutative
$\U(1)$ instanton moduli spaces described by the hyper-K\"ahler
quotient construction with $N=1$. Geometrically, as described in
Ref.~\cite{MY}, the critical-point sets
are precisely the fixed submanifolds of the $\U(1)^{N-1}$ tri-holomorphic
Killing vectors on $\widehat\ms_{k,N}^{(\zeta)}$ defined by the VEVs.

\section{The Centred One-Instanton Partition Function}

We now use the localization technique to evaluate the centred one-instanton
partition function.

The instanton effective action has $N$
critical points, labelled by $v\in\{1,2,\ldots,N\}$, at which \eqref{critp}
\EQ{
\Bchi=-\Bphi_v\ ,\qquad w_{u\aD}\propto\delta_{uv}\ .
\label{critpn}
}
Note that $a'_n=0$ in the one-instanton sector.
Without-loss-of-generality, we choose our 
non-commutativity parameters 
\EQ{
\zeta^1=\zeta^2=0\ ,\qquad\zeta^3\equiv\zeta>0\ .
\label{ncom}
}
In this case the ADHM constraints
\eqref{adhm} are solved with
\EQ{
w_{u\aD}=\sqrt{\zeta}e^{i\theta}\delta_{uv}\delta_{\aD1}\ ,
}
for an arbitrary phase angle $\theta$. The integrals over $w_{v\aD}$ are
then partially saturated by the three $\delta$-functions in
\eqref{bmes} that impose the ADHM constraints, leaving a trivial
integral over the phase angle $\theta$:
\EQ{
\int d^2w_v\,d^2\bar
w_v\,\prod_{c=1}^3\delta\big(\tfrac12\tau^{c\aD}{}_\bD(
\bar w_v^\bD w_{v\aD}-\zeta\delta^{c3})\big)
=8\pi\zeta^{-1}\ .
\label{yuppa}
}
The $\delta$-functions for the Grassmann ADHM constraints in \eqref{intsf}
are saturated
by the integrals over $\{\mu^A_v,\bar\mu^A_v\}$:
\EQ{
\int\,d\mu^A_v\,d\bar\mu^A_v\,\prod_{\aD=1}^2\delta\big(
\bar w_{v\aD}\mu^A_v+w_{v\aD}\bar\mu^A_v\big)=\zeta\ ,
\label{yuppb}
}
for each $A=1,2$.
The remaining variables, $\{w_{u\aD},\mu^A_{u},\bar\mu^A_u\}$, $u\neq
v$, as well as $\{\K_{if},\tilde\K_{fi}\}$, 
are all treated as Gaussian fluctuations around the critical point.
To this order, the instanton effective action \eqref{jslmm} is
\EQ{
S=4\pi^2\bigg\{\zeta\Bchi^2+\sum_{{u=1\atop(\neq v)}}^N
\Big(\Bphi_{vu}^2\big|w_{u\aD}\big|^2
+\tfrac
i{2}\phi_{vu}^\dagger
\bar\mu^A_u\mu_{uA}\Big)
+\tfrac 14\sum_{f=1}^{N_F}(m_f+\phi_v)
\K_f\tilde\K_f\bigg\}+\cdots\ ,
}
where $\Bphi_{uv}\equiv\Bphi_u-\Bphi_v$.
The integrals are easily done. Note that the integral over $\Bchi$
yields a factor of $\zeta^{-1}$ which cancels against the factors of
$\zeta$ arising from \eqref{yuppa} and \eqref{yuppb} so the final
result is, as expected, independent of $\zeta$. Summing over the $N$ 
critical-point sets 
gives the centred one-instanton partition function
\EQ{
\widehat{\EuScript Z}_{1,N}^{\sst(\N=2,N_F)}=
\sum_{v=1}^N\Big\{\prod_{{u=1\atop(\neq
v)}}^N\frac{1}{\phi_{vu}^2}\,
\prod_{f=1}^{N_F}(m_f+\phi_v)\Big\}\ .
\label{oneres}
}
Notice that the resulting expression is holomorphic in the VEVs.

At this point, 
we invite the reader to compare the efficacy of the localization
method compared with the explicit integral performed in 
Ref.~\cite{KMS}. 

\section{The Centred Two-Instanton Partition Function}

We now evaluate the centred two-instanton partition function
using localization. There are two kinds of critical-point sets.
In both cases
\EQ{
\Bchi=-\MAT{\Bphi_{u_1}&0\\ 0&\Bphi_{u_2}}\ ,
}
with $u_1<u_2$, in the first case, giving a critical point set 
\EQ{
\F=\ms_{1,1}^{(\zeta)}\times\ms_{1,1}^{(\zeta)}/{\mathbb
R}^4\simeq{\mathbb R}^4\ ,
\label{case1}
}
where the final form represents the relative position of two points in
${\mathbb R}^4$.
The second case $u_1=u_2$, gives a critical-point set
\EQ{
\F=\widehat\ms_{2,1}^{(\zeta)}\ .
}
We now evaluate these two contribution separately. As in the
one-instanton sector we choose the non-commutativity parameters as in
\eqref{ncom}. 

\subsection{Contribution from 
$\F=\ms_{1,1}^{(\zeta)}\times\ms_{1,1}^{(\zeta)}/{\mathbb R}^4$}

In the first case \eqref{case1}, on the critical-point set \eqref{critp}, the
ADHM constraints are solved with
\EQ{
w_{ui\aD}=\sqrt\zeta e^{i\theta_i}\delta_{uu_i}\delta_{\aD1}
\ ,\qquad
a'_n=\MAT{Y_n&0\\ 0&-Y_n}\ .
\label{diagsol}
}
The two phase angles $\theta_i$, $i=1,2$, are not genuine moduli since they can
be separately rotated by transformations in the $\U(2)$
auxiliary group. The variables $Y_n$ are the genuine moduli
representing the relative positions of the two single
non-commutative $\U(1)$ instantons. 
The corresponding solution of the fermionic ADHM constraints 
\eqref{fadhm} on the critical-point set is
\EQ{
\mu^A=\bar\mu^A=0\ ,\qquad\CM^{\prime A}_\alpha=\MAT{\rho^A_\alpha&0
\\ 0&-\rho^A_\alpha}\ ,
\label{fdiagsol}
}
where $\rho^A_\alpha$ are the superpartners of $Y_n$. 

We now proceed to evaluate the contribution to the 
centred instanton partition function from the critical-point set.
Starting from its linearized form \eqref{ipf}, 
first we expand around the critical values \eqref{diagsol} and
\eqref{fdiagsol}. It is convenient to use the following notation for
the fluctuations
\EQ{
\delta a'_n=\MAT{0&Z_n\\ Z_n^*&0}\ ,\qquad\delta\CM^{\prime A}_\alpha=
\MAT{0&\sigma^A_\alpha
\\ \varepsilon^A_\alpha&0}
} 
and to make the shift
\EQ{
\Bchi\to\Bchi-\MAT{\Bphi_{u_1}&0\\ 0&\Bphi_{u_2}}\ ,
}
so that $\Bchi=0$ on the critical submanifold.
We then integrate over the Lagrange multipliers $D^c$ and
$\bar\psi_A^\aD$ which impose the ADHM constraints \eqref{adhm} and
\eqref{fadhm}. The diagonal components of the constraints (in $i,j$
indices) are the ADHM constraints of the two single $\U(1)$ instantons.
The off-diagonal components vanish on the critical-point set and must
therefore be expanded to linear order in the fluctuations.
For the bosonic variables we have
\AL{
\sqrt\zeta e^{-i\theta_1}(w_{u_12})_2+\sqrt\zeta
e^{i\theta_2}(w_{u_21})_2^*+4i\bar\eta^1_{mn}
Y_mZ_n&=0\ ,\label{ladhm3}\\
-i\sqrt\zeta e^{-i\theta_1}(w_{u_12})_2+i\sqrt\zeta
e^{i\theta_2}(w_{u_21})_2^*+4i\bar\eta^2_{mn}Y_mZ_n&=0\ ,\\
\sqrt\zeta e^{-i\theta_1}(w_{u_12})_1+\sqrt\zeta
e^{i\theta_2}(w_{u_21})_1^*+4i\bar\eta^3_{mn}Y_mZ_n&=0\ ,\label{ladhm1}
}
where $\bar\eta^c_{mn}=\tfrac1{2i}{\rm
tr}_2(\tau^c\bar\sigma_m\sigma_n)$ are 't~Hooft's
$\eta$-symbols. Similarly in the Grassmann sector
\AL{
\sqrt\zeta e^{i\theta_2}\bar\mu^A_{1u_2}+2(\rho^{\alpha
A}Z_{\alpha1}-\sigma^{\alpha A}Y_{\alpha1})&=0\ ,\label{oo1}\\
\sqrt\zeta e^{-i\theta_1}\mu^A_{u_12}+2(\rho^{\alpha
A}Z_{\alpha2}-\sigma^{\alpha A}Y_{\alpha2})&=0\ ,\\
\sqrt\zeta e^{i\theta_1}\bar\mu^A_{2u_1}+2(\varepsilon^{\alpha
A}Y_{\alpha1}-\rho^{\alpha A}Z^*_{\alpha1})&=0\ ,\\
\sqrt\zeta e^{-i\theta_2}\mu^A_{u_21}+2(\varepsilon^{\alpha
A}Y_{\alpha2}-\rho^{\alpha A}Z^*_{\alpha2})&=0\ ,
\label{oo4}
}
where $Y_{\alpha\aD}=Y_n\sigma_{\alpha\aD n}$, {\it etc\/}.
These equations correspond to a set of 
linear relations between the fluctuations.
It is convenient to define 
\EQ{
(w_{u_12})_1=e^{i\theta_1}(\xi+\lambda)\ ,\qquad(w_{u_21})_1^*=
e^{-i\theta_2}(-\xi+\lambda)\ ,
}
so that the fluctuations $\xi$ drops out from \eqref{ladhm1}. 
We can use \eqref{ladhm3}-\eqref{ladhm1} to solve for 
$(w_{u_12})_2$, $(w_{u_21})_2$ and $\lambda$, and \eqref{oo1}-\eqref{oo4} to
solve for
$\mu^A_{1u_2}$, $\mu^A_{2u_1}$, $\bar\mu^A_{u_12}$ and $\bar\mu^A_{u_21}$. 
We then
use the $\U(2)$ symmetry to fix (i) the fluctuation $Z_n$ to be orthogonal
to $Y_n$, $Z_nY_n=0$; and (ii) $\theta_i=0$. The Jacobian for
the first part of this gauge fixing is
\EQ{
\frac1{{\rm Vol}\,\U(2)}\int d^{12}a'\to\frac{16}{\pi^2}
\int d^4Y\,d^3Z\,d^3Z^*\,Y^2\ .
}

Now we turn to expanding 
the instanton effective action \eqref{jslmm}. First the bosonic
pieces. To Gaussian order around the critical point
\EQ{
S_{\rm b}=S^{(1)}_{\rm b}+S^{(2)}_{\rm b}+\cdots\ ,
}
where
\EQ{
\frac1{4\pi^2}S^{(1)}_{\rm
b}=\zeta\big(\Bchi_{11}^2+\Bchi_{22}^2\big)+8Y^2
|\Bchi_{12}|^2+2\big|\Bphi_{u_1u_2}\xi+\sqrt\zeta\Bchi_{12}
\Big|^2
+2\Bphi_{u_1u_2}^2\big(1+4\zeta^{-1}Y^2\big)|Z|^2
\label{act1}
}
and 
\EQ{
\frac1{4\pi^2}S^{(2)}_{\rm
b}=\sum_{i=1}^2
\sum_{{u=1\atop(\neq u_1,u_2)}}^N
\Bphi_{uu_i}^2\big|w_{ui\aD}\big|^2\ .
} 
In order to simplify the integration over the fluctuations, it is
convenient to shift
\EQ{
\xi\to
\xi-\sqrt\zeta\frac{\Bphi_{u_1u_2}\cdot\Bchi_{12}}{\Bphi_{u_1u_2}^2}
}
and define the orthogonal decomposition
\EQ{
\Bchi=\Bchi^\parallel+\Bchi^\perp\ ,\qquad
\Bchi^\perp\cdot\Bphi_{u_1u_2}=0\ .
}
After having done this
\eqref{act1} becomes
\SP{
\frac1{4\pi^2}S^{(1)}_{\rm
b}&=\zeta\big(\Bchi_{11}^2+\Bchi_{22}^2\big)+2\zeta\big(1+4\zeta^{-1}Y^2\big)
|\Bchi_{12}^\perp|^2+8Y^2
\big|\Bchi_{12}^\parallel\big|^2
\\
&+2\Bphi_{u_1u_2}^2\Big(|\xi|^2+\big(1+4\zeta^{-1}Y^2\big)|Z|^2\Big)
} 

To Gaussian order the Grassmann parts of the 
instanton effective action \eqref{jslmm} are
\EQ{
S_{\rm f}=S_{\rm f}^{(1)}+S_{\rm f}^{(2)}+\cdots\ ,
}
where
\SP{
\frac1{4\pi^2}S_{\rm f}^{(1)}&=
-\tfrac i2\phi^\dagger_{u_1u_2}\big(1+4\zeta^{-1}Y^2\big)
\sigma^{\alpha A}\varepsilon_{\alpha A}+
i\rho^{\alpha A}\big(2\phi_{u_1u_2}^\dagger
Z_{\alpha\aD}\bar
Y^{\aD\beta}+\chi_{12}^\dagger\delta_\alpha{}^\beta\big)
\varepsilon_{\beta A}\\
&+i\sigma^{\alpha A}\big(2\phi_{u_1u_2}^\dagger
Y_{\alpha\aD}\bar
Z^{*\aD\beta}-\chi_{21}^\dagger\delta_\alpha{}^\beta\big)\rho_{\beta A}
-2i\zeta^{-1}\phi_{u_1u_2}^\dagger\rho^{\alpha A}Z_{\alpha\aD}\bar
Z^{*\aD\beta}\rho_{\beta A}
}
and 
\EQ{
\frac1{4\pi^2}S_{\rm f}^{(2)}=
\tfrac i2\sum_{i=1}^2\sum_{{u=1\atop(\neq u_1,u_2)}}^N
\phi^\dagger_{uu_i}\bar\mu^A_{iu}\mu_{uiA}
+\tfrac14\sum_{i=1}^2\sum_{f=1}^{N_F}
(m_f+\phi_{u_i})\K_{if}\tilde\K_{fi}\ .
}
By shifting the fluctuations $\sigma^A$ and $\varepsilon^A$ by the 
appropriate amounts of $\rho^A$, we can complete the square yielding 
\SP{
\frac1{4\pi^2}S^{(1)}_{\rm f}&=
-\tfrac i2\phi^\dagger_{u_1u_2}\big(1+4\zeta^{-1}Y^2\big)
\sigma^{\alpha A}\varepsilon_{\alpha A}\\
&-2i\zeta^{-1}(1+4\zeta^{-1}Y^2)^{-1}\rho^{\alpha
A}\Big(\phi^\dagger_{u_1u_2}
Z_{\alpha\aD}\bar Z^{*\aD\beta}+2\chi_{12}^\dagger
Y_{\alpha\aD}\bar Z^{*\aD\beta}+2\chi_{21}^\dagger Z_{\alpha\aD}
\bar Y^{\aD\beta}\Big)\rho_{\beta A}\ .
}

Before we proceed, let us remind ourselves that only the variables $Y_n$
and $\rho^A_\alpha$ are facets of the critical-point set, the
remaining variables are all fluctuations.
The contribution to the centred 
instanton partition function from the critical-point set is then
proportional to
\SP{
&
\int d^4Y\,d\xi\,d\xi^*\,d^3Z\,d^3Z^*\,d^{8}\Bchi\,
d^4\rho\,d^4\sigma\,d^4\varepsilon\\ &\times\ 
\prod_{i=1}^2\bigg\{\prod_{u=1\atop(\neq u_1,u_2)}^Nd^2w_{ui}\,
d^2\bar w_{iu}\,d^2\mu_{ui}\,d^2\bar\mu_{iu}\,
\prod_{f=1}^{N_F}d\K_{if}\,d\tilde\K_{fi}\bigg\}\ Y^2\ 
\exp(-S_{\rm b}^{(1)}-S_{\rm b}^{(2)}-S_{\rm f}^{(1)}-S_{\rm f}^{(2)})\
.
}
The integrals over the Grassmann variables 
$\{\sigma^A_\alpha,\varepsilon^A_\alpha,\rho^A_\alpha\}$ are saturated by 
pulling down terms from $S_{\rm
f}^{(1)}$ yielding the factors
\EQ{
(\phi_{u_1u_2}^\dagger)^4\zeta^{-2}(1+4\zeta^{-1}Y^2)^2\Big(4Y^2(\chi_{21}^\dagger
Z-\chi_{12}^\dagger Z^*)^2+(\phi^\dagger_{12})^2(Z^2Z^{*2}-(Z\cdot
Z^*)^2)\Big)\ .
}
The integrals over the remaining Grassmann variables
$\{\mu_{ui}^A,\bar\mu_{iu}^A,\K_{if},\tilde\K_{fi}\}$, $u\neq u_1,u_2$,
are saturated 
by pulling down terms from $S_{\rm f}^{(2)}$ giving rise to
\EQ{
\prod_{i=1}^2\prod_{{u=1\atop(\neq
u_1,u_2)}}^N(\phi^\dagger_{uu_i})^2\prod_{f=1}^{N_F}
(m_f+\phi_{u_1})(m_f+\phi_{u_2})\ .
}

The $\{Z,\xi,\Bchi\}$ integrals are
\SP{
&\int d\xi\,d\xi^*\,
d^3Z\,d^3Z^*\,d^8\Bchi\,\Big(4Y^2(\chi_{21}^\dagger
Z-\chi_{12}^\dagger Z^*)^2+(\phi^\dagger_{u_1u_2})^2(Z^2Z^{*2}-(Z\cdot
Z^*)^2)\Big)\,e^{-S^{(1)}_b}\,\\ 
&=2^{-21}3\pi^{-8}\frac{(\phi^\dagger_{u_1u_2})^2}
{\zeta^3\Bphi_{u_1u_2}^{12}Y^2(1+4\zeta^{-1}Y^2)^6}
}
while those over $w_{ui\aD}$, $u\neq u_1,u_2$, give a factor
\EQ{
\prod_{i=1}^2\prod_{u=1\atop(\neq
u_1,u_2)}^N\frac1{\Bphi_{uu_i}^4}\ .
}

Finally all that remains is to integrate over the relative position of
the instantons:
\EQ{
\int d^4Y\,\frac{1}{\zeta^2(1+4\zeta^{-1}Y^2)^4}=\frac{\pi^2}{92}\ .
}
Putting all the pieces together with the correct numerical factors
gives the final contribution of the critical-point set to the
centred instanton partition function
\EQ{
\frac2{\phi_{u_1u_2}^6}\prod_{i=1}^2\prod_{{u=1\atop(\neq
u_1,u_2)}}^N\frac1{\phi_{uu_i}^2}\prod_{f=1}^{N_F}
(m_f+\phi_{u_1})(m_f+\phi_{u_2})\ .
}
Notice that the result is holomorphic in the VEVs as required.
Summing over the $\tfrac12N(N-1)$ critical-point sets of this type gives
the following contribution
\EQ{
\sum_{{u,v=1\atop
(u\neq v)}}^N\frac{S_u(\phi_u)S_v(\phi_v)}{\phi_{uv}^2}\ ,
\label{res2}
}
where we have written the answer in terms of the functions
\EQ{
S_u(x)\equiv \prod_{{v=1\atop(\neq u)}}^N
\frac1{(x-\phi_v)^2}\prod_{f=1}^{N_F}(m_f+x)
\label{defs}
}
defined in \cite{D'Hoker:1997nv}.

\subsection{Contribution from $\F=\widehat\ms_{2,1}^{(\zeta)}$}

There are $N$ critical-point sets of this type for which $u_1=u_2\equiv
v\in\{1,\ldots,N\}$. On the critical submanifold  $\{w_{vi\aD},a'_n\}$
and $\{\mu^A_{vi},\bar\mu^A_{iv},\CM^{\prime
A}_\alpha\}$ satisfy the ADHM constraints, \eqref{adhm} and
\eqref{fadhm}, respectively, of two
instantons in a non-commutative $\U(1)$ theory. The remaining
variables all vanish and are treated as fluctuations around the
critical-point set.

As previously, it is convenient to
shift the auxiliary variable $\Bchi$ by its critical-point value:
\EQ{
\Bchi\to\Bchi-\Bphi_v1_{\sst[2]\times[2]}\ .
\label{shift}
}
We now expand in the fluctuations $\{w_{ui\aD},\mu^A_{ui},
\bar\mu^A_{iu}\}$, for $u\neq v$, 
as well as $\{\K_{if},\tilde\K_{fi}\}$. Since all the components of the
ADHM constraints are non-trivial at leading order the fluctuations
decouple from the $\delta$-functions in \eqref{bmes} and \eqref{intsf}
which impose the constraints. The fluctuation integrals only involve
the integrand $\exp-S$, where $S$ is expanded to Gaussian order around
the critical-point set.
However, it is important, as we shall see below, to leave $\Bchi$
arbitrary rather than set it to its critical-point value; namely,
$\Bchi=0$, after the shift \eqref{shift}. The fluctuation 
integrals produce the non-trivial factor
\EQ{
\prod_{u=1\atop(\neq v)}^N
\frac1{\big({\rm det}_2(\chi+\phi_{uv}1_{\sst[2]\times[2]})\big)^2}
\prod_{f=1}^{N_F}\det_2\big((m_f+\phi_v)1_{\sst[2]\times[2]}-\chi\big)
=S_v(\phi_v-\lambda_1)S_v(\phi_v-\lambda_2)\ .
\label{above}
}
Here, $\lambda_i$, $i=1,2$, are the eigenvalues of the $2\times2$
matrix $\chi$ and $S_u(x)$ was defined in \eqref{defs}.

The remaining integrals involve the supersymmetric volume integral on
$\widehat\ms_{2,1}^{(\zeta)}$, into which we insert the integrand
\eqref{above} which depends non-trivially on $\chi$. Now by itself 
$\int_{\widehat\ms_{2,1}^{(\zeta)}}\Bomega^{\sst(\N=2)}=0$.
This is clear from the linearized form \eqref{ipf}:
integrals over the Grassmann collective coordinates pull down two elements
of the matrix $\chi^\dagger$ from the 
action and since there are no compensating factors of $\chi$ the resulting
integrals over the phases of the elements of 
$\chi$ will integrate to zero. This is why we
left $\chi$ arbitrary in \eqref{above} since after expanding in powers
of the eigenvalues $\lambda_i$ it is potentially the quadratic terms 
that will give a non-zero result when inserted into
$\int_{\widehat\ms_{2,1}^{(\zeta)}}\Bomega^{\sst(\N=2)}$. To quadratic order
\eqref{above} is
\EQ{
\tfrac12S_v(\phi_v)\frac{\partial^2S_v(\phi_v)}{\partial\phi_v^2}
(\lambda_1^2+\lambda_2^2)+\PD{S_v(\phi_v)}{\phi_v}\PD{S_v(\phi_v)}{\phi_v}
\lambda_1\lambda_2\ .
}
So the contribution from
this critical-point set is of the form
\EQ{
{\EuScript I}_1S_v(\phi_v)\frac{\partial^2S_v(\phi_v)}{\partial\phi_v^2}
+{\EuScript I}_2\PD{S_v(\phi_v)}{\phi_v}\PD{S_v(\phi_v)}{\phi_v}\ ,
\label{ress}
}
where the VEV-independent constants ${\EuScript I}_{1,2}$ are given by the  
following integrals (in the linearized form \eqref{ipf})
\SP{
&{\EuScript
I}_1=\tfrac12\int_{\widehat\ms_{2,1}^{(\zeta)}}\Bomega^{\sst(\N=2)}
\,(\lambda_1^2+\lambda_2^2)
\equiv\int_{\widehat\ms_{2,1}^{(\zeta)}}\Bomega^{\sst(\N=2)}\,\big(
\tfrac12({\rm tr}_2\chi)^2-{\rm det}_2\chi\big)\ ,\\
&{\EuScript
I}_2=\int_{\widehat\ms_{2,1}^{(\zeta)}}\Bomega^{\sst(\N=2)}\,\lambda_1\lambda_2
\equiv\int_{\widehat\ms_{2,1}^{(\zeta)}}\Bomega^{\sst(\N=2)}
\,{\rm det}_2\chi\ .
\label{defi}
}
We remark that \eqref{ress} is holomorphic in the VEVs as required. 

The moduli space $\widehat\ms_{2,1}^{(\zeta)}$ is the
Eguchi-Hanson manifold \cite{Lee:2001hp}, a well-known four-dimensional
hyper-K\"ahler space. So after all the Grassmann variables 
and $\Bchi$ have been integrated out, 
we can write ${\EuScript I}_{1,2}$ as integrals over the
Eguchi-Hanson space of a suitable integrand.
The Appendix is devoted to proving 
\EQ{
{\EuScript I}_1=\frac14\ ,\qquad{\EuScript I}_2=0\ .
}
Hence the final result for the contributions from the $N$ critical
points of this type to the centred instanton partition function is
\EQ{
\tfrac14\sum_{u=1}^N
S_u(\phi_u)\frac{\partial^2S_u(\phi_u)}{\partial\phi_u^2}\ .
\label{res1}
}

Finally, summing \eqref{res1} and \eqref{res2} we have the
centred two-instanton partition function
\SP{
\widehat{\EuScript Z}_{2,N}^{\sst(\N=2,N_F)}=
\sum_{{u,v=1\atop(u\neq v)}}^N
\frac{S_u(\phi_u)S_v(\phi_v)}{\phi_{uv}^2}+
\tfrac14\sum_{u=1}^NS_u(\phi_u)\frac{\partial^2S_u(\phi_u)}
{\partial\phi_u^2}\ .
}

\section{The One- and Two-Instanton Contributions to the Prepotential}

Now that we have calculated the centred one- and two-instanton
partition functions using localization we can proceed to write down
the one- and two-instanton contributions to the prepotential using
\eqref{ndeq}. Writing the expression in terms of the quantity $S_u(x)$
defined in \eqref{defs}, we have
\AL{
{\cal F}_1^{\text{nc}}&=\sum_{u=1}^NS_u(\phi_u)\ ,
\label{onei}\\
{\cal F}_2^{\text{nc}}&=
\sum_{u,v=1\atop(u\neq v)}^N\frac{S_u(\phi_u)S_v(\phi_v)}{\phi_{uv}^2}+
\tfrac14\sum_{u=1}^NS_u(\phi_u)\frac{\partial^2S_u(\phi_u)}
{\partial\phi_u^2}\ ,
\label{twoi}
}
where the label reminds us that the result is really calculated in the
non-commutative theory.

The one-instanton contribution \eqref{onei} should be compared with the
brute-force calculation of the one-instanton contribution reported in
Ref.~\cite{KMS}\footnote{In order to compare with this reference our VEVs
should be multiplied by $-\sqrt2$.}
\EQ{
{\cal F}_1=
\prod_{u,v=1\atop(u\neq
v)}^N\frac1{(\phi_v-\phi_u)^2}\,\prod_{f=1}^{N_F}
(m_f+\phi_v)\Big\}+S_1^{\sst(N_F)}\ .
\label{KMSr}
}
The extra contribution compared with \eqref{onei} is
\SP{
&S_1^{\sst(N_F<2N-2)}=0\ ,\quad S_1^{\sst(2N-2)}=
-2^{2(1-N)}\MAT{2N-3\\ N-1}\ ,\quad
S_1^{\sst(2N-1)}=-2^{2(1-N)}\MAT{2N-3\\ N-1}\sum_{f=1}^{N_F}m_f\ ,\\
&S_1^{\sst(2N)}=-2^{2(1-N)}\MAT{2N-3\\ N-1}
\sum_{f,f'=1\atop(f<f')}^{N_F}m_fm_{f'}-2^{-2N}\MAT{2N\\
N-1}\sum_{u=1}^N\phi_u^2\ .
\label{exc}
}
There is an interesting interpretation of this 
extra contribution to ${\cal F}_1$, denoted ${\cal F}_\partial$
in Ref.~\cite{KMS}. The brute-force calculation in
Ref.~\cite{KMS} involved the commutative theory and so an integral
over $\widehat\ms_{1,N}$. For $k=1$, and $\zeta^c=0$, the ADHM
constraints \eqref{adhm} are solved with 
\EQ{
w
=\rho\,\Omega
\begin{pmatrix}1_{\sst [2]\times [2]}\\
0_{\sst [N-2] \times [2]} \end{pmatrix} \ ,
\label{wcon2}
}
where $\Omega\in\SU(N)$ and we think of $w$ as a $N\times2$ matrix
with elements $w_{u\aD}$. The parameter $\rho$ is identified with the
instanton scale size. Taking into account the stabilizer of the group
action as well as the quotient by $\U(1)$ transformations
\eqref{symm}, the moduli space has the form of a cone:
\EQ{
\widehat\ms_{1,N}\simeq{\mathbb
R}^+\times\frac{\SU(N)}{\text{S}\big(\U(N-2)\times\U(1)\big)}
\ ,
}
where $\rho$ is the coordinate along the cone. This space has conical
a singularity at the apex of the cone $\rho=0$ 
where the instanton has shrunk to zero size. It is this singularity
with is smoothly resolved in the
non-commutative theory. Interestingly the extra contribution denoted
${\cal F}_\partial$ in Ref.\cite{KMS} 
actually arises from the singularity itself.
To see this, one has to follow in detail the analysis of
Ref.~\cite{KMS}. The strategy followed is to perform the integral in
the linearized form \eqref{ipf} and leave the $\chi$ (denoted $z$ in
Ref.~\cite{KMS}) integral until
last. The final $\chi$ integrals can then be performed using Gauss'
Theorem. There are pole contributions which give rise to the 
first term in \eqref{KMSr} and a boundary contribution from the circle
at infinity in $\chi$-space. It is this latter contribution which
gives rise to ${\cal F}_\partial$. However, this region of
$\chi$-space corresponds to the singularity of $\widehat\ms_{1,N}$ as
can be seen by the following argument. Consider again the linearized
integral form for the partition function \eqref{ipf}. Suppose we
integrate out the variable $\chi$ by solving its equation-of-motion.
This gives $|\chi|\sim\rho^{-1}$ so that the large circle in
$\Bchi$-space corresponds to $\rho=0$.
In the non-commutative theory the singularity is resolved and the
contribution $S_1^{\sst(N_F)}$ disappears. In fact using the formalism
of \cite{KMS} with the addition of the non-commutativity parameters $\zeta^c$
one can show (as was done for the case $N_F=0$ in
\cite{Hollowood:2001ng}) that the contribution from the circle at
infinity in $\Bchi$-space vanishes.
So we have a nice intuitive picture of the difference between the
calculations in the commutative and non-commutative theories: the
latter misses a contribution from the singularity corresponding to
point-like instantons. 

Now we compare our results to the Seiberg-Witten curves
proposed in Refs.~\cite{Klemm:1995qs,Argyres:1995xh}, 
for $N_F=0$, and
Refs.~\cite{Hanany:1995na,Argyres:1995wt,Minahan:1996er}, 
for $1\leq N_F<2N$ (restricted to $N\leq3$ for the last
reference). The finite theory with $N_F=2N$ will be considered
separately. Extracting even the one-instanton coefficient of the
prepotential is a lengthy calculation undertaken in 
Refs.~\cite{D'Hoker:1997nv,Ito:1996qj}. It is interesting that the
Seiberg-Witten theory predictions precisely match the expression
\eqref{onei} of the non-commutative theory. In other words, the
predictions do not re-produce the additional contribution
$S^{\sst(N_F)}_1$ which, as we have argued, arises from the singularity of
the moduli space. However, for $N_F<2N$ this contribution is only a
constant (VEV-independent) contribution to the prepotential 
which is entirely unphysical since the low-evergy effective
action only depends on derivatives of the prepotential.
 Now we turn to the two-instanton sector with $N_F<2N$. If the picture at the
one-instanton level is matched at the two-instanton level then we
should expect \eqref{twoi} to match the prediction from Seiberg-Witten
theory. The two-instanton contribution was extracted from the curve
via a lengthy calculation in
Ref.~\cite{D'Hoker:1997nv}. It is astonishing to find exact agreement
with our calculation of ${\cal F}_2$ using localization \eqref{twoi}.
Hence, the main conclusion we can draw is that this is
strong evidence in favour of the hypothesis made in
Ref.\cite{Hollowood:2001ng} that the Seiberg-Witten predictions for
the prepotential in the commutative and non-commutative theories are
identical.

The case with $N_F=2N$ is rather special since the underlying theory
in this case is finite. There are two issues here; namely the 
relation of the, on the one hand, commutative, and on the other,
non-commutative, instanton calculations, to the predictions from the
Seiberg-Witten curves. The relation of the commutative instanton
calculations to the prediction of the Seiberg-Witten curve has already
been extensively studied
\cite{MO-I,MO-II,Dorey:1997bn,Argyres:2000ty}. In particular, we
believe that a general picture has been established in the latter
reference. Of particular importance is the notion of
re-parameterizations of the couplings, masses and VEVs. Since it is
known that conventional commutative instanton calculations are
consistent with the curves, we find it more convenient to simply
compare the commutative and non-commutative instanton calculations.
We find that for identical infra-red physics the 
the commutative and non-commutative theories should be formulated in
terms of two different coupling constants, $\tau$ and
$\tau_{\text{nc}}$ with
\EQ{
\tau_{\text{nc}}=\tau+\sum_{k=1}^\infty c_ke^{2\pi ik\tau}\ ,
}
where the $c_k$ are constants. For the commutative 
theory with $N=2N_F$ the prepotential has the semi-classical 
expansion 
\EQ{
{\cal F}\Big|_{N_F=2N}=\tfrac12\tau\sum_{u=1}^N\phi^2_u+
\frac1{2\pi i}\sum_{k=1}^\infty{\cal F}_ke^{2\pi ik\tau}\ .
}
The expansion is identical in the non-commutative theory with $\tau$
replaced by $\tau_{\text{nc}}$. By the hypothesis of
\cite{Hollowood:2001ng}, the prepotentials in both cases can only
differ by a physically irrelevant VEV-independent piece, hence
\AL{
{\cal F}_1&={\cal F}_1^{\text{nc}}+i\pi c_1\sum_{u=1}^N\phi_u^2\
,\label{comp1}\\
{\cal F}_2&={\cal F}_2^{\text{nc}}+2\pi ic_1{\cal
F}_1^{\text{nc}}+i\pi c_2
\sum_{u=1}^N\phi_u^2\ ,\label{comp2}
}
{\it etc\/}, modulo VEV-independent constants. Comparing our result 
\eqref{onei} with the brute-force calculation in Ref.~\cite{KMS} gives
\EQ{
c_1=-\frac1{2^{2N}i\pi}\MAT{2N\\
N-1}\ .
}
For the case $\SU(2)$ we can also
compare \eqref{twoi} 
with the explicit brute-force integration over the two
instanton moduli space described in
Ref.~\cite{MO-II}. One then finds
\EQ{
c_2\Big|_{\SU(2)}=\frac1{2^3\pi
i}\Big(1+\frac7{2^43^5}-\frac{13}{2^4}\Big)\ .
}
Note that the consistency of \eqref{comp1} and \eqref{comp2} 
is non-trivial because the coefficients ${\cal F}_{1,2}$ and ${\cal
F}_{1,2}^{\text{nc}}$ all depend non-trivially on the masses.
Notice that consistency at the two-instanton level is maintained by
assuming that the masses in commutative and non-commutative theories are equal.

What about the situation for instanton number $k>2$? It is clear that
the localization technique can be extended beyond $k>2$. The answer
for ${\cal F}_k$ will then be written as a sum over the contributions
from the different critical-point sets \eqref{gfps} and each
contribution will involve an integral over the product of 
$\U(1)$ instanton moduli spaces.
The holy grail of this quest would be to reproduce
the recursion relations of 
Refs.~\cite{Edelstein:1999sp,Edelstein:1999dd,Edelstein:2000xk,Chan:2000gj} 
for the coefficients ${\cal F}_k$ and thereby completely prove Seiberg-Witten
theory using conventional semi-classical field theory methods.

Finally we should mention that these localization techniques are also
applicable to calculating the instanton coefficients 
of the prepotential of the $\N=2$ theory that
arises from a mass-deformation of the $\N=4$ theory. This situation is
considered in Ref.~\cite{MY}.

I would to thank my colleagues Nick Dorey, Valya Khoze and Prem Kumar
for many discussions about these matters.

\startappendix

\Appendix{Integrals on the Eguchi-Hanson Manifold}

The space $\widehat\ms_{2,1}^{(\zeta)}$ is the Eguchi-Hanson manifold
as can be seen by explicitly solving the ADHM constraints. Here we
follow the treatment in Ref.~\cite{Lee:2001hp}. Up to (most
of) the $\U(2)$ symmetry
\EQ{
w_1=\MAT{\sqrt{1-b}&\sqrt{1+b}}\ ,\qquad w_2=0\ ,\qquad
a'_n=y_n\MAT{1&\sqrt{\tfrac{2b}{a}}\\ 0&-1}\ ,
}
where
\EQ{
a=\tfrac12\zeta^{-1}y^2\ ,\qquad b=\frac1{a+\sqrt{1+a^2}}\ .
\label{defab}
}
Defining coordinates
\EQ{
z_0\equiv y_2-iy_1=r\cos\tfrac\theta2 e^{i(\psi+\varphi)/2}\ ,\qquad
z_1\equiv y_4-iy_3=r\sin\tfrac\theta2 e^{i(\psi-\varphi)/2}\ ,
}
we find
\EQ{
\int_{\widehat\ms_{2,1}^{(\zeta)}}\Bomega=2\pi^2\int
r^3dr\,\sin\theta\,d\theta\,d\varphi\,d\psi\ ,
}
where $0\leq\theta\leq\pi$, $0\leq\varphi,\psi\leq2\pi$.

The solution of the fermionic ADHM constraints can be written as 
\SP{
\CM^{\prime1A}&=\MAT{\sqrt{\tfrac a{2b}}(z_0\sigma^A+z_1\varepsilon^A)&
z_0\sigma^A\\ z_1\varepsilon^A& -\sqrt{\tfrac
a{2b}}(z_0\sigma^A+z_1\varepsilon^A)}\ ,\\
\CM^{\prime2A}&=\MAT{\sqrt{\tfrac a{2b}}(z_1^*\sigma^A-z_0^*\varepsilon^A)&
z_1^*\sigma^A\\ -z_0^*\varepsilon^A& -\sqrt{\tfrac
a{2b}}(z_1^*\sigma^A-z_0^*\varepsilon^A)}\ ,
}
along with $\mu^A=\bar\mu^A=0$. The Grassmann part of the integration
measure \eqref{intsf} is 
\EQ{
\pi^{-8}
\frac{r^4b^8}{(1-b^2)^4(1+b^2)^2}\int d^2\sigma\,d^2\varepsilon\ .
}
The centred partition function, written using the auxiliary variables
$\Bchi$ \eqref{ipf}, is then
\EQ{
\widehat{\EuScript Z}_{2,1}=2^4\int
r^3dr\,d^3\Omega\,d^2\sigma\,d^2\varepsilon\,d^8\Bchi\,\frac{r^4b^6}{
(1-b)^4(1+b)^2(1+b^2)}\,e^{-S}\ .
}
The instanton effective action is
\EQ{
S=4\pi^2\chi\BL\chi^\dagger
-2i\pi^2r^2\varepsilon^A\sigma_A\big(\chi^\dagger_{11}
-\sqrt{\tfrac{2a}b}\chi_{12}^\dagger-\sqrt{\tfrac{2a}b}
\chi_{21}^\dagger-\chi^\dagger_{22}\big)\ ,
}
where
\EQ{
\BL=\MAT{1&0&0&-b\\ 0&0&1+2a+b&\sqrt{1-b^2}\\
0&1+2a+b&0&\sqrt{1-b^2}\\
-b&\sqrt{1-b^2}&\sqrt{1-b^2}&1+2b}
}
in the basis where $\chi$ is to thought of as the 4-vector
$\chi=(\chi_{11},\chi_{12},\chi_{21},\chi_{22})$.

Integrating over the Grassmann variables gives
\SP{
&\widehat{\EuScript Z}_{2,1}=2^7\pi^4\int
r^3dr\,\sin\theta\,d\theta\,d\varphi\,d\psi\,d^8\Bchi\\
&\times\,\frac{r^8b^6}{
(1-b)^4(1+b)^2(1+b^2)}\,\Big(\chi^\dagger_{11}
-\sqrt{\tfrac{2a}b}\chi_{12}^\dagger-\sqrt{\tfrac{2a}b}\chi_{21}^\dagger
-\chi^\dagger_{22}
\Big)^2\exp(-4\pi^2\chi\BL\chi^\dagger)\ .
}
The integral over the phases of $\chi$ clearly vanish and so 
$\widehat{\EuScript Z}_{2,1}=0$. 

Now we consider the two integrals defined in \eqref{defi}. Inserting
the general factor $F(\chi_{ij})$, we can
perform the integrals over $\Bchi$ and the angles to arrive at
\SP{
&\int
r^3dr\,\frac{r^8b^8}{
(1-b^2)^4(1+b^2)^2}\\
&\times\ \Big(\PD{}{J_{11}}
-\sqrt{\tfrac{2a}b}\PD{}{J_{12}}-\sqrt{\tfrac{2a}b}
\PD{}{J_{21}}-\PD{}{J_{22}}
\Big)^2\ F\Big(\PD{}{J_{ij}^\dagger}\Big)\ e^{
J\BL^{-1}J^\dagger}\Big|_{J=0}\ .
}
Using this general formula one finds
\EQ{
{\EuScript I}_1=4\int_0^\infty dr
\frac{r^{11}b^8}{(1-b^4)^4}=\frac14\ ,\qquad
{\EuScript I}_2=0\ ,
}
where $b$ is a function of $r$ given in \eqref{defab}.

\end{document}